%% file: main.tex
\documentclass[apj,twocolumn]{openjournal}

\usepackage{lipsum}

\usepackage{xcolor}
\usepackage{textgreek}
\usepackage[utf8]{inputenc}
\usepackage[english]{babel}
\usepackage{hyperref}	
\hypersetup{colorlinks=true,linkcolor=blue,citecolor=blue,filecolor=blue,urlcolor=blue}

\usepackage{color,colortbl}
\usepackage{tensind}
\tensordelimiter{?}
\DeclareGraphicsExtensions{.bmp,.png,.jpg,.pdf}
\usepackage{verbatim}
\usepackage[normalem]{ulem}
\usepackage{orcidlink}
\usepackage{soul}

\usepackage{newtxtext,newtxmath}
\usepackage{graphicx}	
\usepackage{amsmath}	
\input{defs}

\usepackage{aas_macros} 

\begin{document}
\title[Light Echoes and SNe Distances]{Echo Location: Distances to Galactic Supernovae From ASAS-SN Light Echoes and 3D Dust Maps\vspace{-15mm}}

\author{Kyle~D.~Neumann \orcidlink{0000-0002-2701-8433},$^{1*}$
Michael~A.~Tucker \orcidlink{0000-0002-2471-8442},$^{2,3}$
Christopher~S.~Kochanek,$^{2,3}$
Benjamin~J.~Shappee \orcidlink{0000-0003-4631-1149},$^{4}$
and K.~Z.~Stanek \orcidlink{0009-0001-1470-8400}$^{2,3}$}
\thanks{$^*$E-mail: kdn5172@psu.edu}

\affiliation{$^{1}$Department of Astronomy and Astrophysics, Pennsylvania State University, 525 Davey Lab, University Park, PA 16802, USA}
\affiliation{$^{2}$Department of Astronomy, The Ohio State University, 140 West 18th Avenue, Columbus, OH 43210, USA}
\affiliation{$^{3}$Center for Cosmology and AstroParticle Physics (CCAPP), The Ohio State University, 191 W. Woodruff Ave., Columbus, OH 43210, USA}
\affiliation{$^{4}$Institute for Astronomy, University of Hawai'i, 2680 Woodlawn Drive, Honolulu, HI 96822, USA}

\date{Accepted 2024 October 24. Received 2024 October 24; in original form 2024 July 19}


\begin{abstract}
    Light echoes occur when light from a luminous transient is scattered by dust back into our line of sight with a time delay due to the extra propagation distance. We introduce a novel approach to estimating the distance to a source by combining light echoes with recent three-dimensional dust maps. We identify light echoes from the historical supernovae Cassiopeia A and SN 1572 (Tycho) in nearly a decade of imaging from the All-Sky Automated Survey for Supernovae (ASAS-SN). Using these light echoes, we find distances of $3.6\pm0.1$~kpc and $3.2^{+0.1}_{-0.2}$~kpc to Cas A and Tycho, respectively, which are generally consistent with previous estimates but are more precise. These distance uncertainties are primarily dominated by the low distance resolution of the 3D dust maps, which will likely improve in the future. The candidate single degenerate explosion donor stars B and G in Tycho are clearly foreground stars. Finally, the inferred reddening towards each SN agrees well with the intervening H~\textsc{i} column density estimates from X-ray analyses of the remnants.
\end{abstract}


\maketitle

\section{Introduction}
\label{sec:intro}

Scattered light echoes (hereafter light echoes or echoes) occur when optical light from a luminous transient scatters off interstellar dust towards the observer. Dust heated by absorption of the radiation can also produce a thermal echo in the mid-infrared (e.g., \citealt{krause05}). The most famous local examples of scattered light echoes are SN~1987A \citep{crotts88} and V838~Mon \citep{bond03, tylenda03}, but \citet{Rest2012} also detected echoes from the Great Eruption of $\eta$ Carinae, \citet[][hereafter \citetalias{rest08b}]{rest08b} detected echoes from the Galactic supernovae (SNe) Cassiopeia A (hereafter Cas A) and SN 1572 (hereafter Tycho or Tycho's SN), and \citet{rest05} detected echoes from several SNe in the Magellanic Clouds. Light echoes can be resolved for SNe in nearby galaxies (e.g., SN 1993J, \citealt{sugerman02}; SN 2014J, \citealt{crotts15,yang17}), but the small angular sizes are typically only resolvable from space. 

\citet{krause08a} classified Cas~A as a Type~IIb SN using spectroscopy of its echoes and \citet{rest11b} subsequently used spectra of multiple echoes along different sight lines to show that Cas A had distinct explosion asymmetries. Similarly, light echo spectra from Tycho's SN \citep{krause08b} confirmed its classification as a thermonuclear (Type~Ia) explosion based on historical records of its light curve \citep{ruiz-lapuente04a}. The spectra of echoes from $\eta$ Carinae provide a unique probe of this enigmatic transient (\citealt{Rest2012}, \citealt{Smith2018}). This shows how light echoes can provide unique probes of historical events long after they have faded.

\begin{figure*}

\begin{center}
    \includegraphics[width=.9\linewidth]{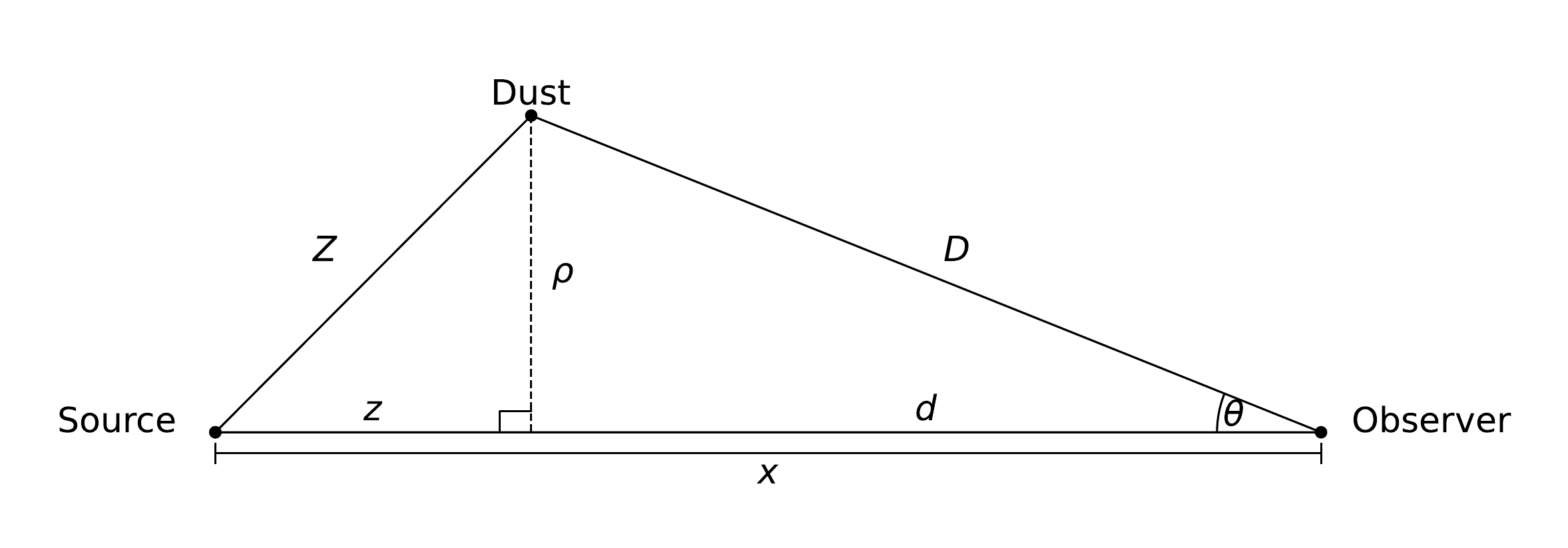}
    \caption{The geometry of a light echo.}
    \label{fig:le_geo}
\end{center}
\end{figure*}

Fig.~\ref{fig:le_geo} shows the geometry of a light echo. \citet{chevalier86}, \citet{sugerman03}, and \citet{tylenda03} provide the basic models for the appearance of a light echo for various dust geometries, and \citet{rest05} expands upon these models for Galactic SNe. The location and motion of echoes determine the distance between the reflecting dust and the source but not the distance to the source or the dust \citep{bond09}. With the addition of polarization measurements, it is possible to determine the overall distance to the source because the maximum polarization constrains the scattering angle to provide the extra geometric constraint needed to make a distance estimate \citep{sparks94}. Versions of this method have been used to estimate the distance to V838 Mon \citep{sparks08} and SN~1987A \citep[e.g.,][]{cikota23}. 

Distances to supernova remnants (SNR) are generally quite uncertain. Only two have direct parallaxes to a pulsar remnant (Vela, \citealt{Dodson2003}, and the Crab, \citealt{Chatterjee2009}) and three are interacting binaries with parallaxes to the companion stars \citep[see ][]{kochanek21}. Dispersion measure models can be used for pulsars, but they are only accurate to $\sim 25\%$ (e.g., \citealt{Taylor1993}). H~\textsc{i} (or CO) radio absorption velocities combined with a Galactic kinematic model can provide approximate distances for SNRs towards the inner galaxy (e.g., first by \citealt{Ilovaisky1972}) and the correlation between radio diameter and surface brightness (the $\Sigma$-$D$ relation) also provides approximate distances (e.g., \citealt{Case1998}). Kinematic models of the SNR expansion, matching proper motions and velocities or simply the expansion rate and size, depend on models for the three-dimensional shape of the expansion or the deceleration history (e.g., \citealt{chevalier80}, \citealt{reed95}, \citealt{hayato10}, \citealt{alarie14} for Cas~A and Tycho). All of these non-parallex methods have significant distance uncertainties leading to substantial uncertainties in remnant masses and energetics. This also leads to difficulties in identifying companion stars unbound in the explosion \citep[e.g.,][for Tycho]{RuizLapuente2004,Kerzendorf2013,shappee13b}. One new approach is to use wide-field, multi-fiber spectrographs to determine the distance at which stars show high velocity Ca~\textsc{ii} or Na~\textsc{i} absorption lines from the SNR (\citealt{kochanek24}).

Recent years have seen the rapid development of three-dimensional (3D) maps of the Galactic dust distribution. While there are older models \citep[e.g.,][]{drimmel03,berry12,green15}, the availability of \textit{Gaia} \citep{gaia16,lindegren18,lindegren21} parallax measurements revolutionized our ability to map the Milky Way dust distribution on scales of kpc, leading to many new models \citep[e.g.,][]{green19, dharmawardena24,Edenhofer2024}. Given a 3D dust map, there are two ways to estimate the distance to a SNR. The first approach is to find the distance where the extinction matches an estimate of the extinction of the SNR, usually from the X-ray gas absorption column density (e.g., \citealt{wang20}) and assuming a gas-to-dust ratio. The second approach, which we introduce here, is through dust echoes, where knowing the distance to the dust supplies the missing information needed to determine the distance from an echo (see Fig.~\ref{fig:le_geo}). This second approach has the advantage that multiple, independent distance measurements can be used if there
are multiple, well-separated echoes.


Using wide-field imaging from the All-Sky Automated Survey for Supernovae (ASAS-SN\footnote{\url{http://www.astronomy.ohio-state.edu/~assassin/}}; \citealt{shappee14}) from 2014 through 2021, we identify light echoes from two galactic SNe, Cas~A and Tycho, and use them to estimate their distances. In our fiducial analysis, we combine the ASAS-SN echoes with the Tycho and Cas~A echoes from \citetalias{rest08b}. We use the {\tt Bayestar19} 3D dust maps (\citealt{green19}). While the \citet{dharmawardena24} maps provide better distance resolution, their distance range ($<2.8$~kpc) is too restricted for these SNRs. We did find that the \cite{green19} and \cite{dharmawardena24} maps agree well inside this distance limit for our lines of sight. In Section~\ref{sec:data}, we describe how we find light echoes in the ASAS-SN data. In Section~\ref{sec:analysis}, we use these 3D dust maps and light echo geometry to constrain the distances to these SN. In Section~\ref{sec:conclusion}, we summarize our results.

\section{The Tycho and Cas~A Echoes In ASAS-SN}
\label{sec:data}

\asassn observes the entire sky with 20 telescopes on 5 mounts (since late-2017) using a tiling of fields. A field is observed using three 90~second exposures, initially in the $V$-band through late-2018 and now in the $g$-band beginning in late-2017 (\citealt{shappee14}, \citealt{kochanek17}, \citealt{hart23}). There are now $\sim4000$ images available per field collected over roughly 10 years. To search for echoes, we coadd data in ``observing year'' bins (i.e., divided by solar conjunctions) and then use image subtraction to look for changes. We start by assessing the quality of the images in a given \asassn field. Images flagged as low-quality by visual inspection, such as those obtained through moderate cloud cover, are immediately removed. Images with large residuals after kernel convolution are also removed as this indicates the photometric quality of the observation is poor. Then, we discard the 33\% of the remaining images with the highest sky backgrounds. These images are typically obtained during nights with high Moon illumination and/or high-altitude cloud cover. The 33\% threshold is somewhat arbitrary, but experimentation found that discarding $\approx 25-35\%$ of the images produces the best results. After applying these quality cuts, the median number of images per field is $\approx 200$ in $V$-band and $\approx 350$ in $g$-band.

Next, the images are coadded by observing season. We tested different methods for scaling the images during the stacking process (e.g., median, mode, sky value) but in the end, we simply averaged all images to create the yearly stack. This worked well because (1) all \asassn images are obtained with the same exposure time ($90$~s) and (2) the lower-quality images were already discarded in the initial quality assessment. The final $V$- and $g$-band yearly stacked images of the region around Tycho's SN and Cas A have $\sim 60$ and $\sim 90$ images, respectively, with median surface brightness limits of $V\approx 22$~\sbu and $g\approx 24$~\sbu.

\begin{figure*}

\begin{center}
    \includegraphics[width=.95\linewidth]{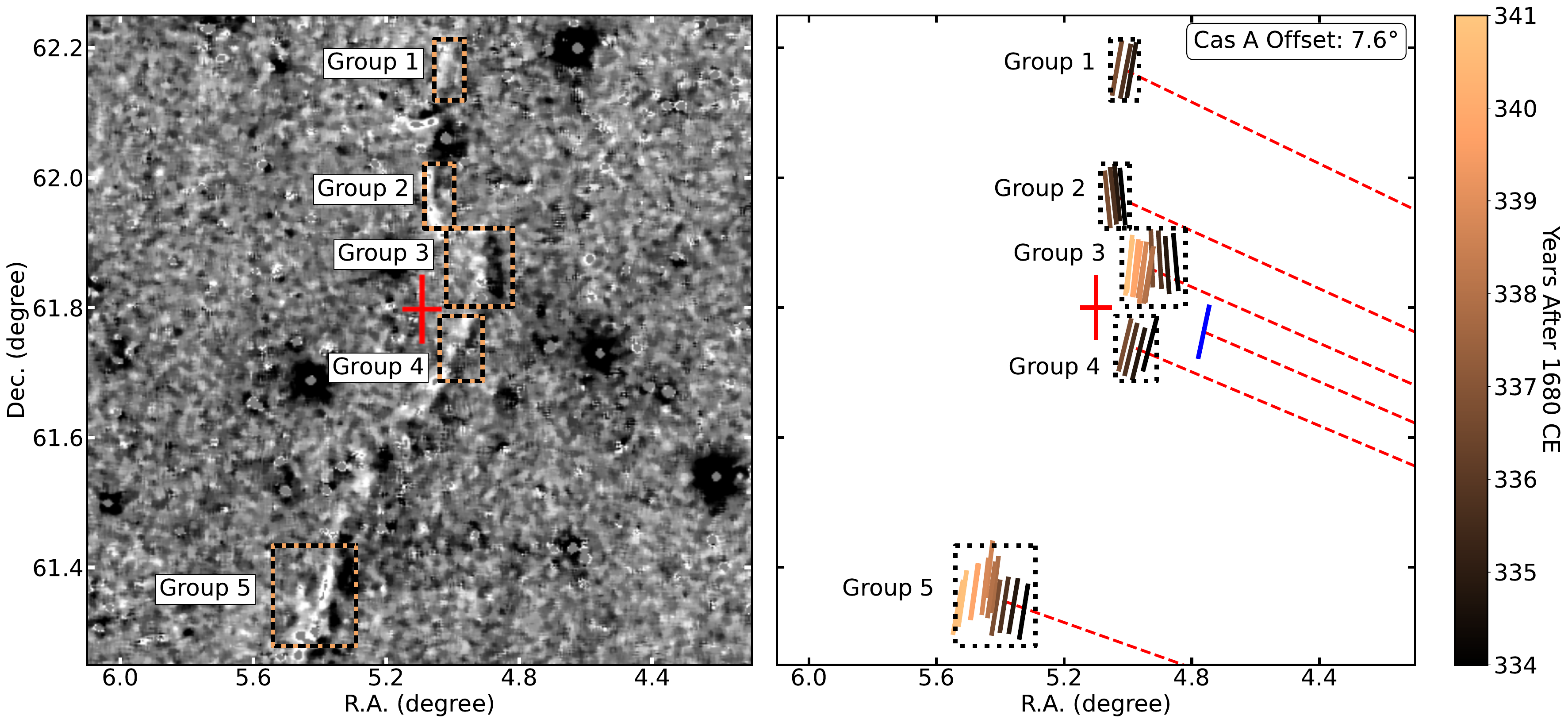}
    \caption{Echoes from Cas~A. The left panel shows the difference image for epochs separated by two years, and the right panel shows all the detected echo groups in this field. The images are very spatially distorted -- the red cross is 0.1\degr by 0.1\degr. The lines show the position and orientation of the echo and color codes the age, showing that the echoes are expanding outwards from Cas~A as expected. Dotted boxes show how the echoes are grouped to deal with the spatial correlations of the dust maps. The blue line is the light echo identified by \citetalias{rest08b}. Dashed red lines extend from each echo group towards the geometric center of the Cas~A remnant. The offset from the red cross to the geometric center of Cas~A is 7.6\degr.}
    \label{fig:cas_echo}
\end{center}
\end{figure*}

Finally, the yearly stacks are pair subtracted using the \textsc{ISIS} image subtraction software \citep{alard2000}. All image pairs are subtracted, so 4 stacked images labeled $1-4$ produce 6 subtraction pairs ($1-2$, $1-3$, $\ldots$, $4-3$). This carries several benefits relative to only subtracting sequential stacks (e.g., $1-2$, $2-3$, $3-4$). Galactic light echoes move tens of arcseconds per year \citepalias{rest08b}, so only subtracting sequential stacks risks self-subtraction of any light echo signal if the angular motion is less than the PSF FWHM ($\approx$ 2 pixels = 16\arcsec). The pair subtraction also helps to separate true light echoes from subtraction artifacts through their time evolution. Subtraction artifacts around bright sources are reduced with an iterative $5\sigma$ clipping routine. Finally, the subtracted images are filtered using a $5\times5$ pixel median filter ($\approx 40\arcsec\times40\arcsec$) to highlight the spatially resolved light echoes and to suppress any remaining subtraction artifacts (see Fig.~\ref{fig:cas_echo}).

\begin{figure}

\begin{center}
    \includegraphics[width=.95\linewidth]{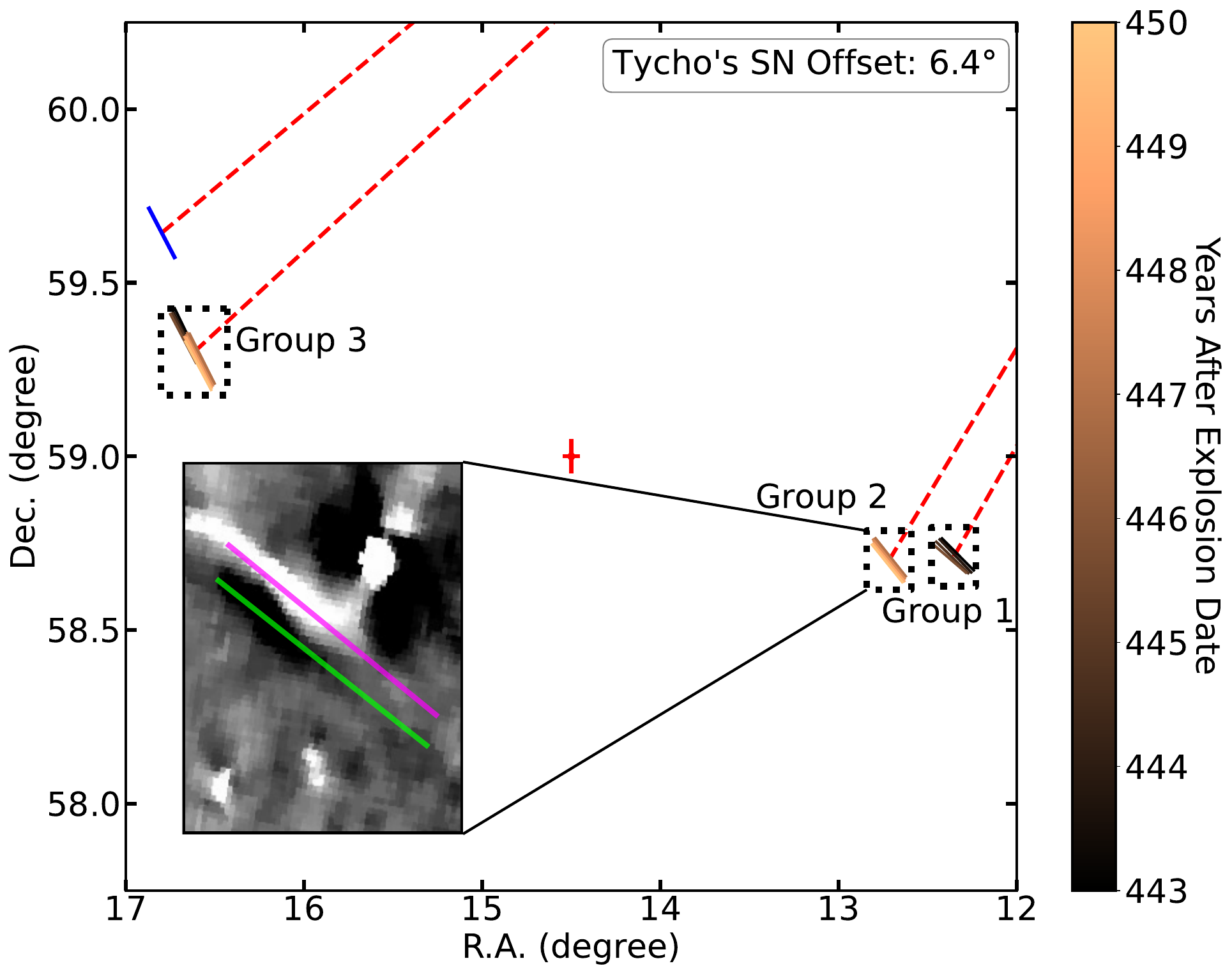}
    \caption{Echoes from Tycho's SN. The zoomed-in image shows the $g$-band Group 2 in images separated by two years where the pink line is the older detection, and the green line is the later detection. These images are significantly distorted --  the red cross is 0.1\degr~by~0.1\degr. The lines show the position and orientation of the echo and the color codes the age, showing that the echoes are expanding outwards from Tycho as expected. Echo groups are identified with boxes and the blue line is the light echo in this field identified by \citetalias{rest08b}. A dashed red line extends from the center of each group towards the center of Tycho. The geometric center of Tycho is 6.4\degr from the red cross.}
    \label{fig:tyc_echo}
\end{center}
\end{figure}

We visually inspected the subtracted images for light echoes, focusing on the region around Cas~A and Tycho's SN and flagging those that show coherence across multiple subtractions. \citet{bhullar21} and \citet{li22} attempted to find light echoes using a machine-learning approach with limited success. While they successfully detected many light echoes, they also had a large number of false positives. For each identified echo, we fit a linear segment to the echo in each subtraction (see Figs.~\ref{fig:cas_echo} and \ref{fig:tyc_echo}), and we average these to calculate their central point and position angle (P.A.). This technique gives us a position with an accuracy of $\sim 20\arcsec$. Nearby echoes with coherent motion away from the SNR are labeled as part of the same echo ``group'' (or simply group). Not all images of a given echo will be used in the subsequent analysis. Some image stacks have fewer images and noisier subtractions or a subtraction artifact interfering with the analysis. We only keep echoes with detections in a least two of the subtracted images. Table~\ref{tab:LE_arc} lists the coordinates, P.A.s, and mean times of observations for the 57 identified echoes.


\section{Analysis} 
\label{sec:analysis}

Creating a dust echo depends on having dust to scatter the light at the distance required by the geometry of Fig.~\ref{fig:le_geo}. The constraint equation for an echo (e.g., \citealt{tylenda03}) is
\begin{equation}\label{eqn:origin}
c \Delta t = (Z - z) + (D - d)
\end{equation}
where 
\begin{equation*}
    \Delta t = t_\mathrm{o} - t_\mathrm{e}
\end{equation*}
is the difference between the time of observation ($t_\mathrm{o}$) and explosion ($t_\mathrm{e}$). The time delay ($c \Delta t$) equals the difference in length between the direct and the dust-scattered paths from the source. If $\theta$ is the observed angle between the echo and the SN, then the distance between the observer and the dust is
\begin{equation}\label{eqn:Deqn}
D = \frac{c\Delta t(c\Delta t+2x)}{2(c\Delta t+x-x\cos{\theta})}.
\end{equation}
For a given source distance, $x$, and explosion time, $t_e$, the uncertainties in the estimate of $D$ are negligible since they only arise from the uncertainties in $t_o$, which is known to a fraction of a year, and $\theta$, which is measured to an accuracy of $\sim 20\arcsec$ out of an overall angle of $\theta \simeq 7^\circ$ (e.g., Figs.~\ref{fig:cas_echo} and \ref{fig:tyc_echo}).

\begin{figure*}

\centering
\includegraphics[width=0.47\linewidth]{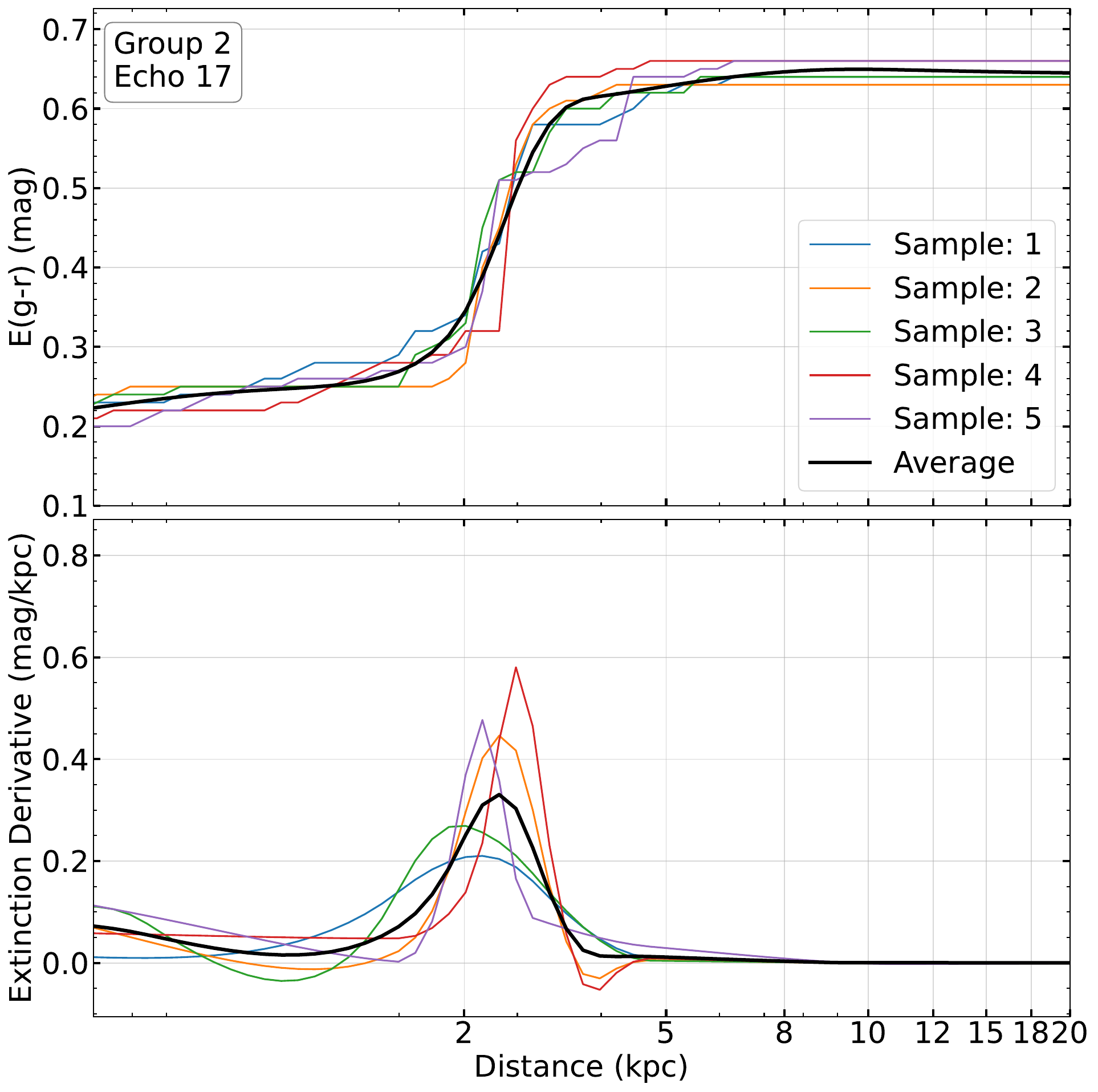}
\includegraphics[width=0.475\linewidth]{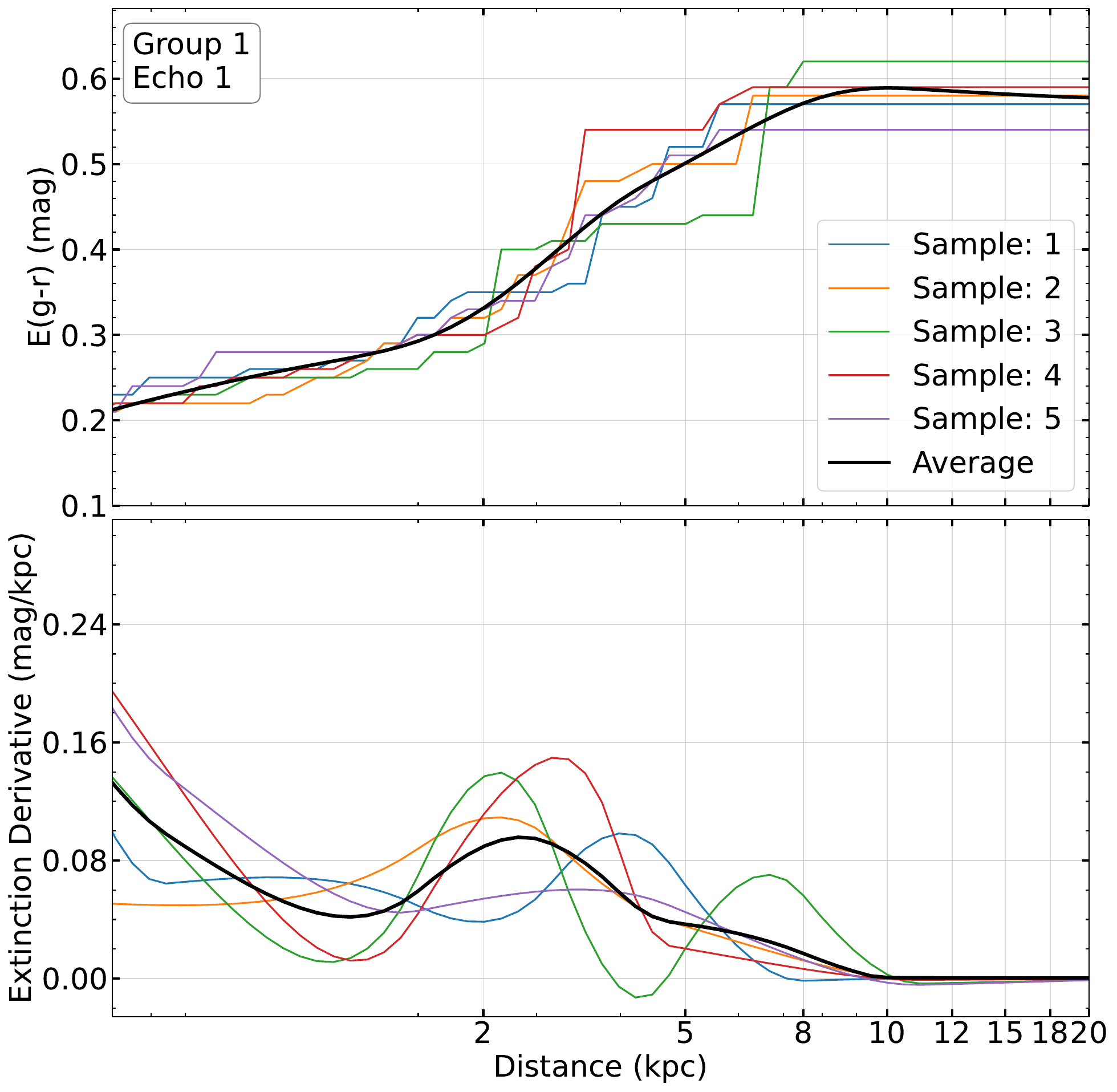}

\caption{Examples of a well (\emph{left}) and a poorly (\emph{right}) localized dust sheet for two Tycho echoes. The upper panels show the integral distributions and the supplied statistical samplings, and the bottom panels show the resulting estimates of the dust density $P(D)$ (colors) and their mean (black). }

\label{fig:dust_geo}
\end{figure*}

Thus, given the echo's location and the time of observation, the observed echo can only be produced for a given source distance and time of explosion if there is dust at a distance $D$ along the line of sight to the echo. The amplitude of the echo is largely determined by the density of dust at that distance with secondary effects from the scattering angle which we will ignore, so we need the dust density as a function of distance $P_j(D_j|x,t_e)$. We fit the {\tt Bayestar19} integral dust distribution with monotonic splines \citep[\texttt{scipy.interpolate.PchipInterpolator},][]{virtanen20} and then use the derivative of the monotonic splines with respect to the distance $D_j$ to define $P_j(D_j|x,t_e)$. Since {\tt Bayestar19} supplies 5 posterior distributions for each sight line, we use the average of the 5 density estimates. Fig.~\ref{fig:dust_geo} shows these distributions for two Tycho echoes with very different dust distributions. In one case, there is a well-defined dust sheet at a distance of 2-3~kpc which will strongly constrain the dust distance, $D_j$. In the second case, the distribution of dust along the line of sight is much broader and will provide little constraint on the dust distance. Because the distance resolution of the {\tt Bayestar19} dust maps is limited, even the thin dust sheet is significantly smeared out in distance. This makes it essentially impossible to usefully constrain the explosion time and limits the precision of the distance estimate from any single echo.

Since the angular resolution of the {\tt Bayestar19} dust maps is $\sim 5 \arcmin$ near Cas~A and Tycho, we will derive essentially the same dust density distribution for echoes with smaller separations. If we multiply the distance posteriors of such echoes, then we are treating their dust distributions as independent measurements, yet they are not. To avoid this, we work in terms of echo groups consisting of all echoes with separations less than $6 \arcmin$. In practice, this means groups form a time sequence of echoes expanding away from the SNR like those seen in Figs.~\ref{fig:cas_echo} and \ref{fig:tyc_echo}. For the echoes $j$ in group $i$ we use the mean of their dust density distributions 
\begin{equation}
    \bar{P}_i(D_i|x, t_e) = \left\langle P_i(D_j|x, t_e) \right\rangle.
     \label{eqn:dustden}
\end{equation}
as a single dust density distribution for the group. This should be a conservative approach to handling the correlations in the dust maps.

The dust distributions, $\bar{P}_i(D_i|x,t_e)$, for each group $i$ should be uncorrelated, so we can multiply their distance posteriors. Using Bayes theorem, our posterior for the distance and age is then 
\begin{equation} 
    P(x, t_e|\hbox{data})\propto P(x) P(t_e) \prod_i \int dD_i \bar{P}_i(D_i | x, t_e)
\label{eqn:bayes}
\end{equation} 
where $P(x)$ and $P(t_e)$ are the priors for the SN distance and explosion time, respectively. Bayesian probability distributions are always normalized to unity, which means that the normalization of the $\bar{P}_i(D_i | x, t_e)$ dust density distributions drops out of the final posteriors. We use a uniform SN distance prior, $P(x)$, from 1.75 kpc to 6.25 kpc for both Cas~A and Tycho. This prior spans their previous distance estimates (see below). We used Gaussian priors for the time of explosion $P(t_e)$ with the parameters defined in Sections \ref{sec:cass} and \ref{sec:tycho}. Using a uniform prior for the time of explosion does not affect the distance estimate. As noted earlier, the low distance resolution of the dust maps means that we are unable to constrain the time of explosion, so the posteriors for the time of explosion are basically just the priors. Fortunately, the distance posterior is essentially uncorrelated with the time of explosion.

As a simple sanity check on the results, we can compare the extinction at the inferred distance to the SNR with the H~\textsc{i} column density $N(H)$ inferred from X-ray observations of the SNR. The $E(g-r)$ extinction of the {\tt Bayestar19} maps can be converted to $E(B-V)=1.11 \times E(g-r)$ (see, e.g., Appendix B in \citealp{green14}), and the visual extinction is related to the column density by
\begin{equation}
    \NH = 8.3\times 10^{21}~\mathrm{cm}^{-2}~\mathrm{mag}^{-1}\times E(B-V)
\end{equation}
\citep[e.g.,][]{liszt2014a, liszt2014b} with a scatter of $\approx 10\%$ \citep[e.g.,][]{lenz2017, nguyen2018, zuo2021}.


\subsection{Cassiopeia A}
\label{sec:cass}

Cas~A is the remnant of a Type IIb~SN \citep{krause08a}, and it was the most recent galactic SN that may have been observed. If the explosion was observed, it would have been the sixth magnitude object 3 Cassiopeiae observed by John Flamsteed on 1680 August 16, but its separation of 12\arcmin\xspace from Cas~A makes the identification uncertain \citep{ashworth80,hughes80}. Studies of the proper motions of the ejecta by \citet{thorstensen01} and \citet{fesen06}
estimate explosion times of 1671 - 1680 and $1681\pm19$~years, respectively, which are consistent with Flamsteed's
observation. Based on this, we adopt a Gaussian prior for $P(t_e)$, centered on 1676 CE with a dispersion of 5~yrs. Using a flat prior spanning 1670 - 1680 produces identical results for the distance.

Fig.~\ref{fig:cas_echo} shows an example of a Cas~A light echo at an offset of approximately $7.6\degr$ from the geometric center of 
$\alpha(J2000)=23^h23^m27\fs77, \delta(J2000)=+58\degr48'49\farcs4$ for Cas~A found by \citet{thorstensen01}. This geometric center lies $6\farcs6\pm1\farcs5$ from the neutron star formed in the SN. The total light echo is approximately 1\degr~in length, and we broke it into five separate echo groups (Groups 1 to 5) where Group 3 and Group 5 are the only segments clearly visible in the $g$-band data. This light echo appears to be a continuation of the light echo located in Field No. 3824 of \citetalias{rest08b}, as shown in Fig.~\ref{fig:cas_echo}. 
From 2014 to 2021, we found 35 echoes in 5 echo groups.
For these five echo groups (1 to 5), we found average annual proper motions of $40 \pm 8$, $27 \pm 7$, $44 \pm 3$, $47 \pm 3$, and $56 \pm 3$ arcsec~yr$^{-1}$, respectively. We analyze these echo groups alongside 6 echoes identified by \citetalias{rest08b} as a complimentary sample. These 6 echoes will be treated as 6 individual echo groups consisting of a single echo each.

\begin{figure*}

\begin{center}
    \includegraphics[width=.90\linewidth]{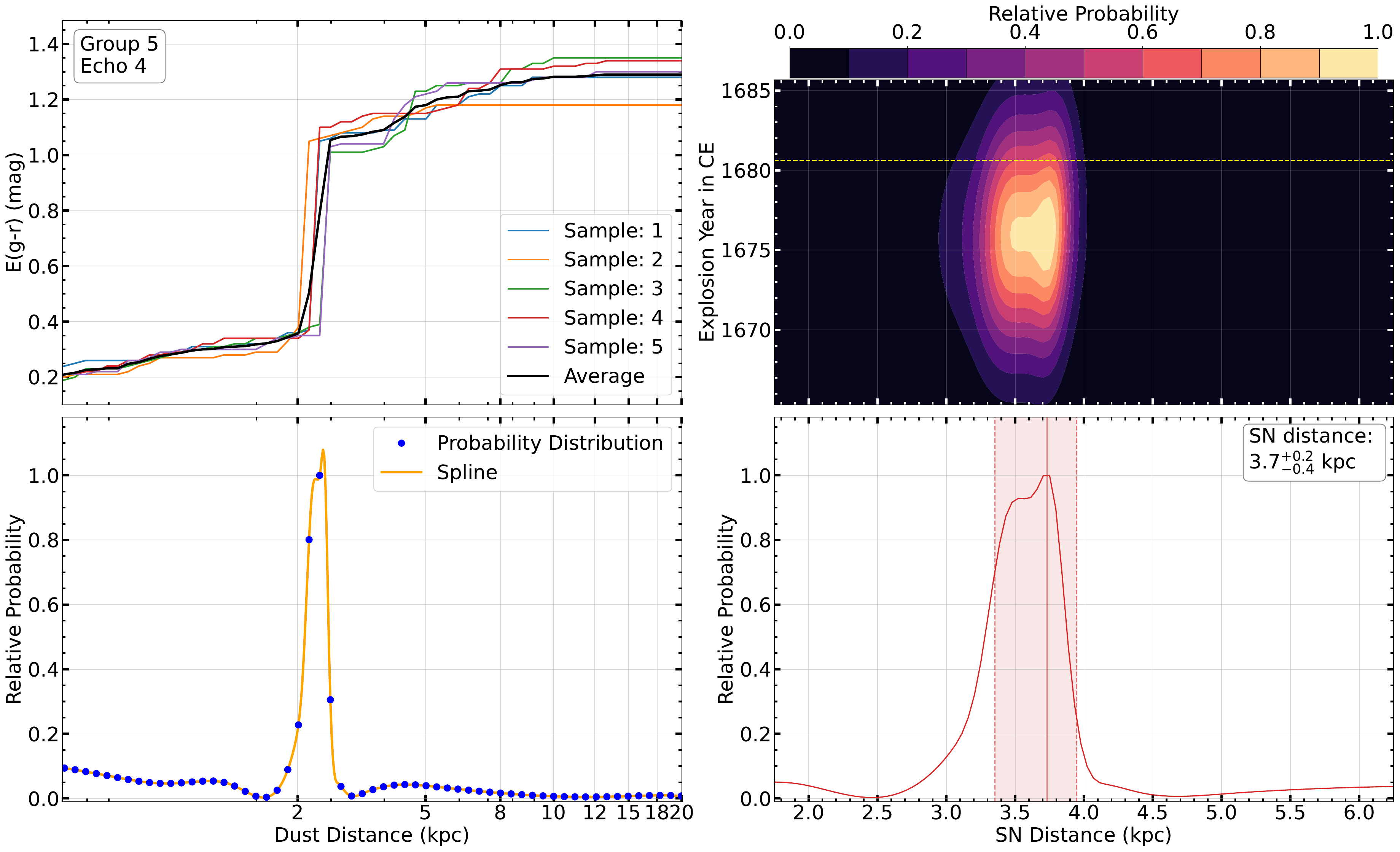}
    \caption{An example of estimating the distance to Cas~A for a single echo (Echo 4 of Group 5). The left panels show the integrated (\emph{top}) and differential (\emph{bottom}) dust distributions. The \emph{upper right} panel shows the 2D probability distribution for the Cas~A distance and its explosion year. The yellow line shows the adopted explosion time of 1680 August 16, when John Flamsteed may have observed Cas~A \citep{ashworth80}. The \emph{lower right} panel is the posterior distribution marginalized along time to show the resulting distance posterior. The solid vertical line is the distance of maximum likelihood, and the dotted lines are the 1$\sigma$ uncertainty.}
    \label{fig:cas_2x2}
\end{center}
\end{figure*}

Fig.~\ref{fig:cas_2x2} shows an example of inferring the Cas A distance from a single echo, Echo 4 in Group 5. Using Eq.~\ref{eqn:Deqn}, the known time of observation, and the offset of the echo from the SNR, we estimate the dust density along the line of 
sight, $P_i(D_i|x,t_e)$ (Eq.~\ref{eqn:dustden}). We integrate over the dust distance weighted by the density (Eq.~\ref{eqn:bayes}) to get a probability distribution for the SN distance and explosion time, $P_i(x,t_e|\mathrm{data})$. We then marginalize the distribution over the time axis to get the posterior
probability distribution for the SN distance, $P_i(x|\mathrm{data})$. As seen in Fig.~\ref{fig:cas_2x2}, a single thin dust sheet dominates this line of sight, leading to a well-constrained distance of $3.7_{-0.4}^{+0.2}$~kpc with a single echo. 
As expected, the posterior for the explosion time is essentially identical to the prior due to the uncertainties associated with dust distances. We also see that the distance is uncorrelated with the explosion time.


\begin{figure}

\begin{center}
    \includegraphics[width=.98\linewidth]{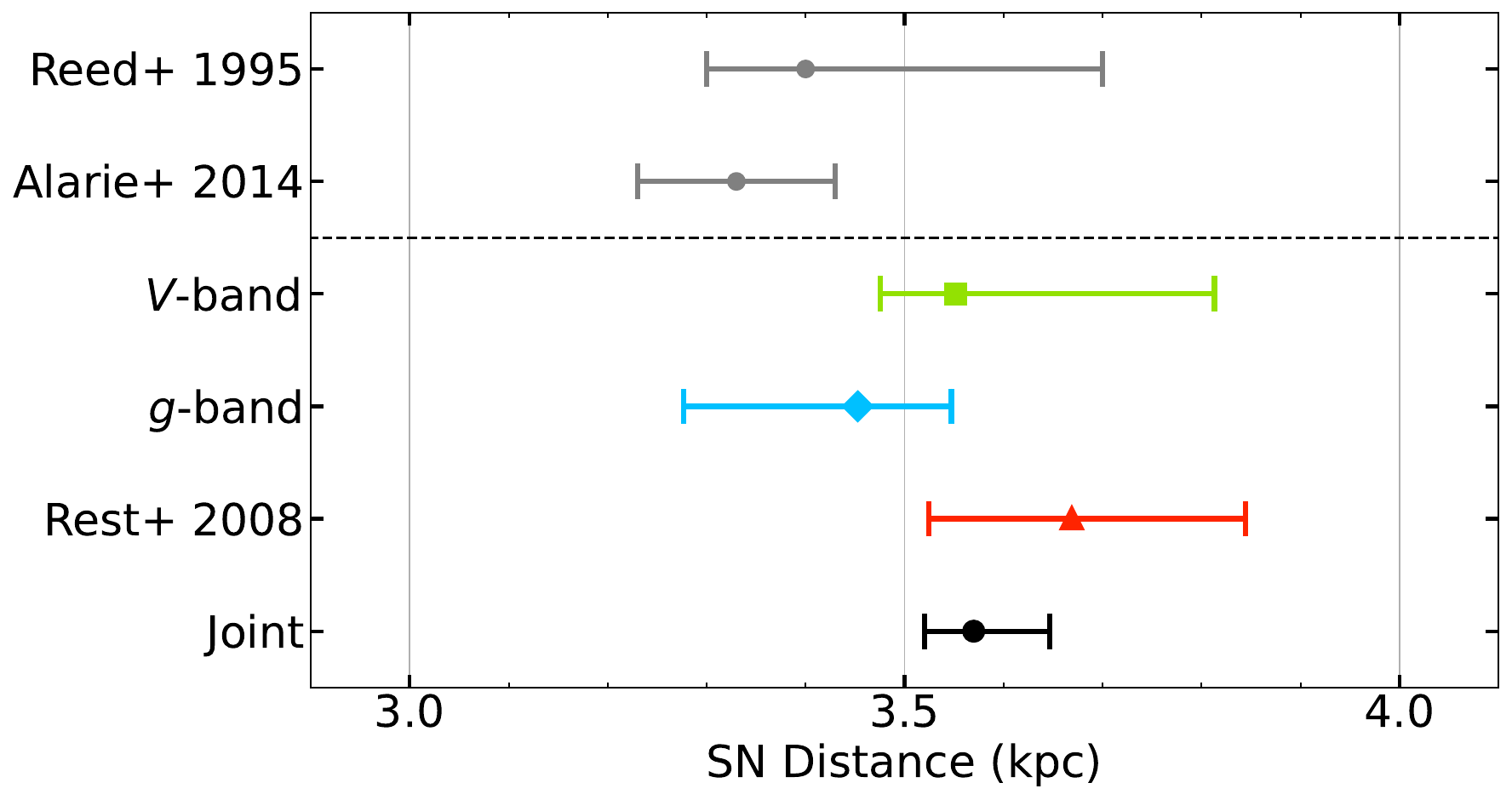}
    \caption{Our estimates for the distance to Cas A (\emph{bottom}, colored) compared to the literature estimates (gray) with 1$\sigma$ uncertainties. Our estimates include ASAS-SN $V$-band echoes (green), ASAS-SN $g$-band echoes (blue), echoes from \citetalias{rest08b} (red), and the joint echo sample (black). The literature distances estimates (\emph{upper}) are from \citet{reed95} and \citet{alarie14}.}
    \label{fig:cas_dist}
\end{center}
\end{figure}

\begin{figure*}

\centering
\includegraphics[width=0.463\linewidth]{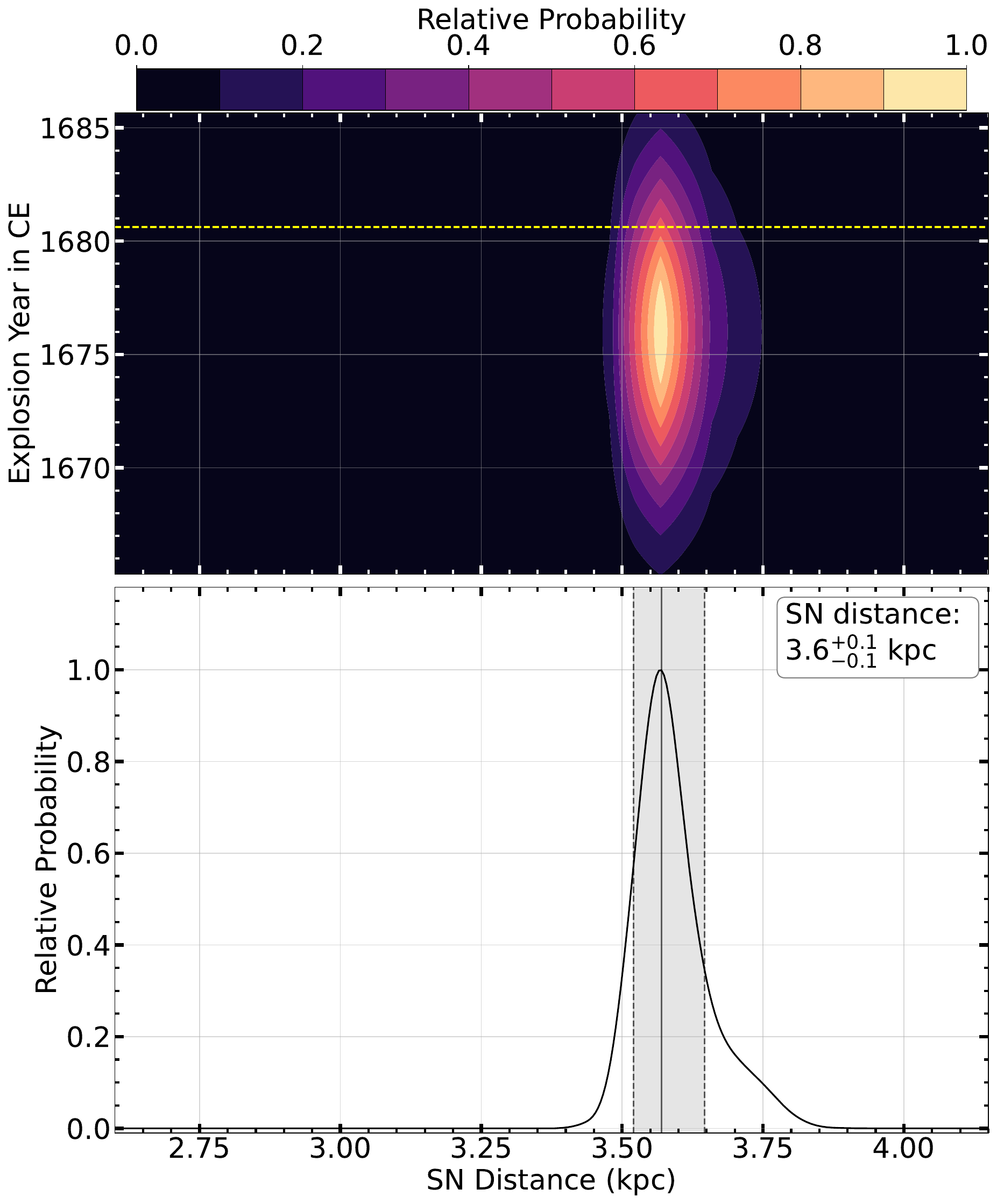}
\includegraphics[width=0.45\linewidth]{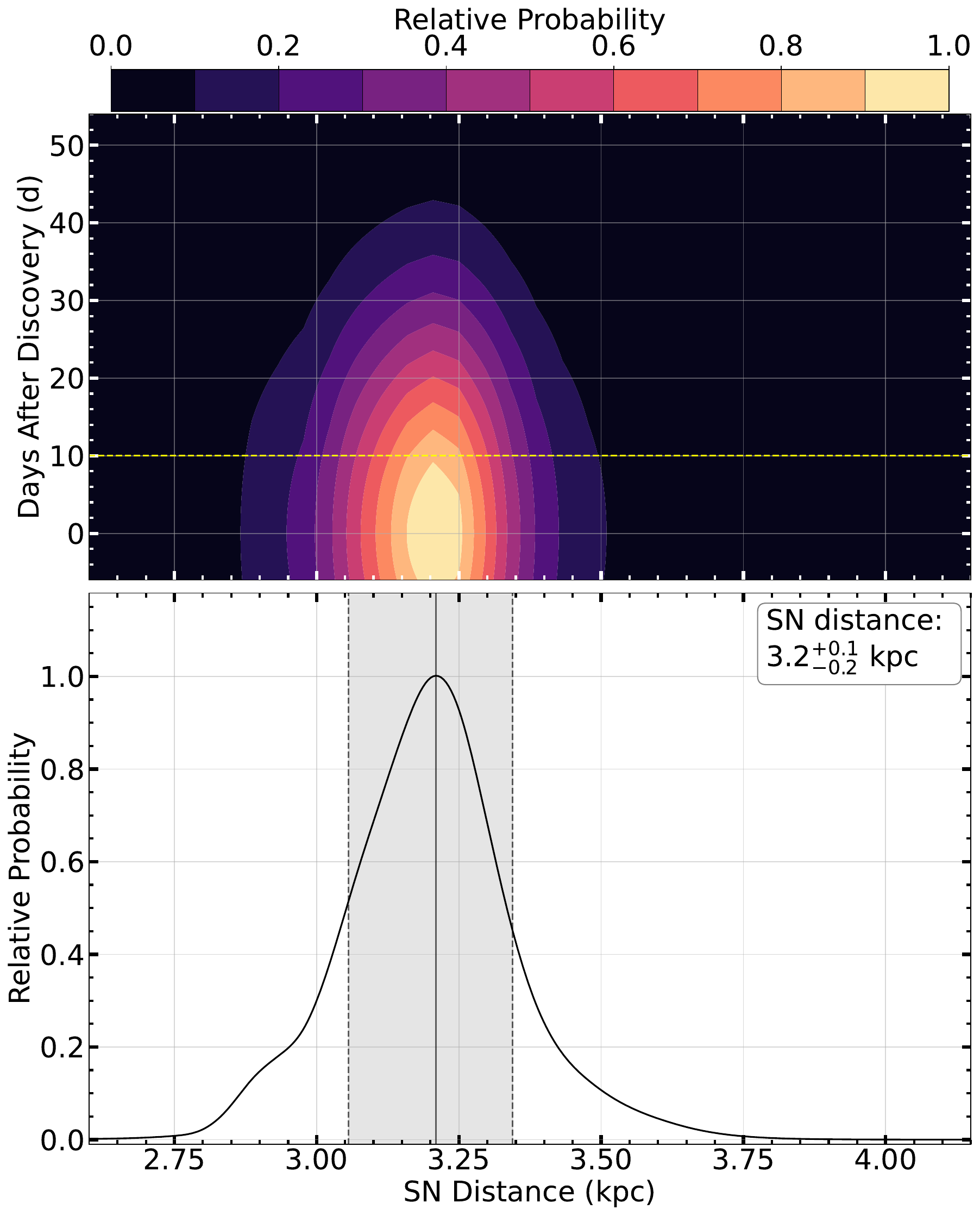}

\caption{2D (\emph{upper}) and explosion time-integrated (\emph{lower}) posterior distributions for the joint samples of Cas A (\emph{left}) and Tycho (\emph{right}) echoes. The yellow dashed lines are the estimated explosion times of 1680 August 16 and 1572 November 21 for Cas A and Tycho, respectively. The solid vertical black lines are the distance of maximum likelihood, and the dotted lines are the 1$\sigma$ uncertainty.}

\label{fig:joint}
\end{figure*}

We did the full calculations for the $V$-band ASAS-SN echo groups, the $g$-band ASAS-SN echo groups, the echoes identified by \citetalias{rest08b}, and the entire joint echo group sample with the results listed in Table~\ref{tab:dist} and shown in Fig.~\ref{fig:cas_dist}. We use the distance at the maximum of the posterior distribution for the central value and the $1\sigma$ uncertainty calculated from the 16\% to 84\% range of the integral posterior probability distributions. All four results are mutually consistent, with a final joint distance estimate of $3.6\pm0.1$~kpc. Fig.~\ref{fig:joint} shows the 2D and explosion time-integrated posterior distributions for this joint sample. Fig.~\ref{fig:cas_dist} also shows other distance estimates for Cas~A. The best existing estimates come from kinematic models of the proper motions and radial velocities of emission line gas in the SNR, with \citet{reed95} finding $3.4^{+0.3}_{-0.1}$~kpc and \citet{alarie14} finding $3.33\pm0.10$~kpc. Our result is consistent with the former and $2\sigma$ inconsistent with the latter. At base, these kinematic models compare the proper motion of the expansion $\mu$ to the line of sight expansion velocity $v$ and determine the distance by matching the two, $d=v/\mu$. While they find the remnant to be fairly spherical, very modest asymmetries could explain the 9\% distance difference with \cite{alarie14}. Even so, such close (9\%) agreement between two radically different methods is encouraging.


Fig.~\ref{fig:SNR_dust} shows the dust distribution along the line of sight to the center of Cas~A along
with our distance estimate. We find that Cas~A lies inside or just behind a dust sheet with significant optical depth.
If we sample this dust distribution over our distance posterior, we find an extinction posterior of
$E(g-r)=1.7_{-0.6}^{+0.3}$. For comparison, \citet{hwang12} measure $N(H)\simeq 2\times10^{22} \mathrm{~cm~}^{-2}$ 
(with no reported uncertainties) from {\it Chandra} X-ray observations, or $E(g-r) \simeq 2.2 \pm 0.3$ if we
assume a 10\% measurement uncertainty and add the 10\% scatter in the extinction to column density conversion.
This is consistent with our result. At least for this sight line, using the dust towards Cas~A to estimate its
distance would give a very strong lower limit because of the dust sheet at $\sim 3.5$~kpc, but a much weaker upper limit 
because the next increase in the integrated extinction lies near 6~kpc and is a fairly weak jump.

\renewcommand{\arraystretch}{1.7}

\begin{table}
\centering
    \caption{Distance Estimates}
    \label{tab:dist}
    \begin{tabular}{c|c|c}
    \hline
        Sample & Cas A~(kpc) & Tycho (kpc) \\
        \hline
        $V$-Band & $3.6^{+0.3}_{-0.1}$ & $4.2^{+0.2}_{-0.9}$ \\
        $g$-Band & $3.5^{+0.1}_{-0.2}$ & $3.3^{+0.2}_{-0.2}$ \\
        Rest+ 08 & $3.7^{+0.2}_{-0.1}$ & $3.1^{+0.2}_{-0.2}$ \\
        Joint & $3.6^{+0.1}_{-0.1}$ & $3.2^{+0.1}_{-0.2}$ \\
        \hline
    \end{tabular}
\end{table}

\subsection{Tycho's Supernova}
\label{sec:tycho}

Tycho's SN was a Type~Ia SN first observed on 1572 November 11 by Jer\'{o}nimo Mu\~{n}oz and Tycho Brahe separately, as discussed by \citet{ruiz-lapuente04a}, and it was observed by European and Asian astronomers for two years. Tycho (SN~1572)
and Kepler (SN~1604) were the most recent, clearly observed SN in the Galaxy given the questions surrounding the possible Flamsteed observation of Cas~A. Spectroscopy of a light echo 400 years later conclusively showed that Tycho's SN was a Type~Ia SN \citep{krause08b}. The direct observations provide a light curve at peak (see \citealt{ruiz-lapuente04a}),
allowing us to use a very tight Gaussian prior on the explosion time centered on the discovery date, 1572 November 11, and with a dispersion of 20 days. Using a flat prior on the explosion time produces identical results.

Fig.~\ref{fig:tyc_echo} shows three echo groups associated with Tycho at an average offset of approximately $6.4\degr$ from the geometric center of $\alpha(J2000)=00^h25^m19\fs9, \delta(J2000)=+64\degr08'18\farcs2$ for Tycho found by
\citet{ruiz-lapuente04b}. Groups 1 and 2 are separated from Group 3 by approximately $4\degr$. Group 3 may be associated with the echo located in field 4821 of \citetalias{rest08b}. Groups 1 and 2 are only seen in the $V$- and $g$-bands, respectively, while Group 3 is seen in both. 
We located 22 echoes associated with 3 echo groups over almost 7 years of observations.
For these three echo groups (1 to 3), we found average annual proper motions of $28 \pm 4$, $28 \pm 4$, and $25 \pm 3$ arcsec~yr$^{-1}$, respectively. 
We analyze these echo groups alongside the 6 echoes identified by \citetalias{rest08b}. Each of these 6 \citetalias{rest08b} echoes are treated as separate echo groups consisting of a single echo.

\begin{figure*}

\begin{center}
    \includegraphics[width=.9\linewidth]{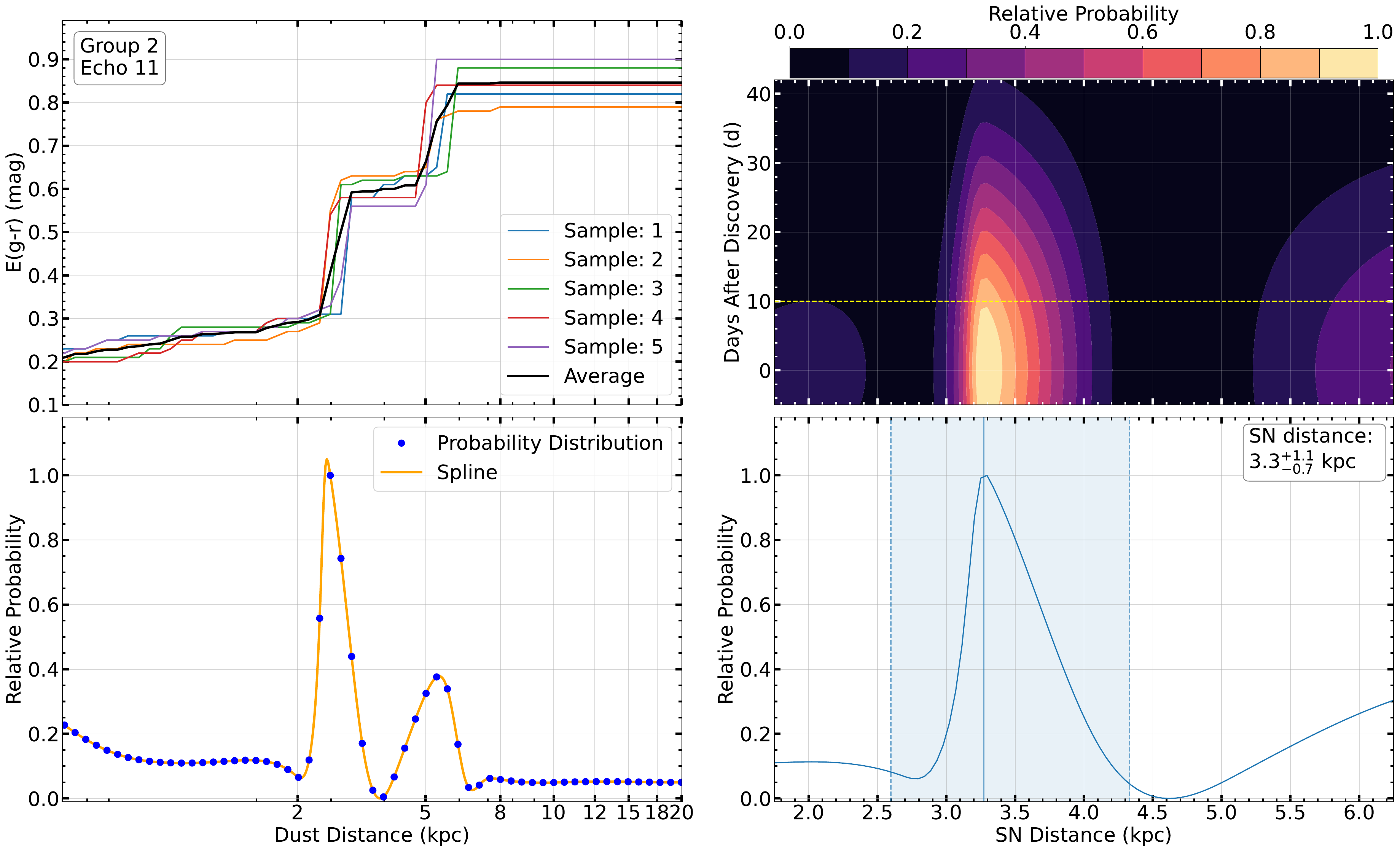}
    \caption{An example of estimating the distance to Tycho's SN for a single echo (Echo 11 of Group 2). The left panels show the integral (\emph{top}) and differential (\emph{bottom}) dust distributions. The \emph{upper right} panel is the 2D probability distribution for the Tycho distance and explosion date where the yellow line is 1572 November 21, the expected day of peak luminosity \citep{ruiz-lapuente04a}. The \emph{lower right} panel is the projected probability distribution for the SN distance based on this echo where the solid vertical line is the distance of maximum likelihood and the dotted lines are the 1$\sigma$ uncertainty.}
    \label{fig:tyc_2x2}
\end{center}
\end{figure*}

\begin{figure}

\begin{center}
    \includegraphics[width=.98\linewidth]{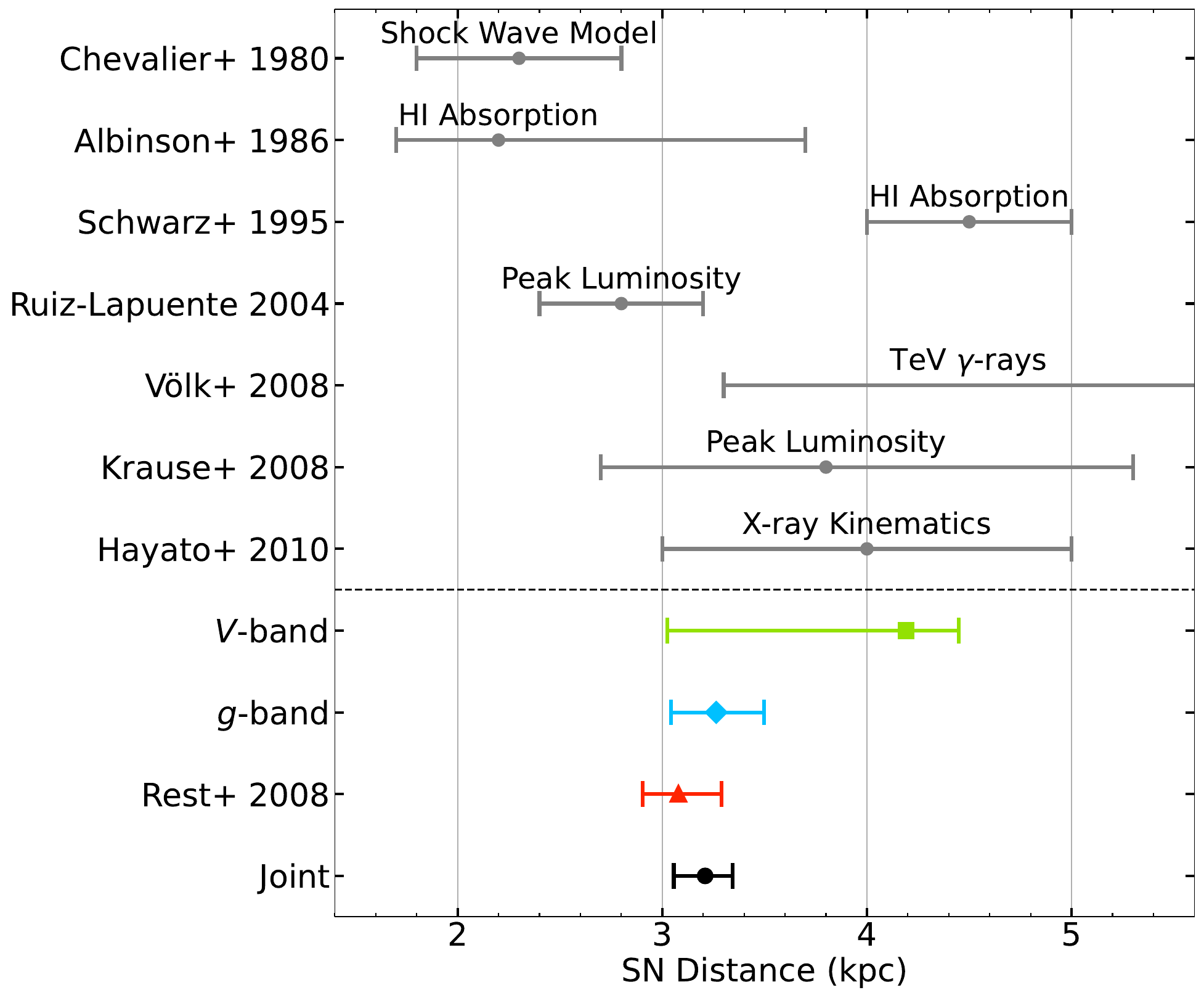}
    \caption{Our estimates for the distance to Tycho's SN (\emph{bottom}, colored) compared to the literature estimates (gray) with 1$\sigma$ uncertainties. Our estimates include the ASAS-SN $V$-band echoes (green), the ASAS-SN $g$-band echoes (blue), the echoes from \citetalias{rest08b} (red), and the joint echo sample (black). The literature distance estimates (\emph{upper}) are from \citet{chevalier80}, \citet{albinson86}, \citet{schwarz95}, 
    \citet{ruiz-lapuente04a}, \citet{volk08}, \citet{krause08b}, \citet{hayato10}.}
    \label{fig:tyc_dist}
\end{center}
\end{figure}

Fig.~\ref{fig:tyc_2x2} is an example of the results for Tycho Echo 13 of Group 2. For this echo, there are two dust sheets along the line of sight, but the second sheet lies close to the 6.25~kpc upper limit of our distance prior. Nonetheless, its presence produces a second probability peak and leads to a posterior distance estimate of $3.3_{-0.7}^{+1.1}$~kpc with a relatively large positive distance uncertainty. This is an example of why having multiple groups helps. As we combine echo groups, the probability associated with the true distance will be reinforced, while the probability associated with such secondary peaks should increasingly cancel provided the distances of any secondary dust sheets are relatively uncorrelated. The explosion time posterior is again unchanged from the prior, as expected, and there is no correlation of the distance estimate with the explosion time. The distance estimates from the $V$-band echo groups, $g$-band echo groups, \citetalias{rest08b} echoes, and the joint echo group sample are given in Table~\ref{tab:dist} and shown in Fig.~\ref{fig:tyc_dist}. The results from the echo samples are mutually consistent, with a final distance estimate of $3.2_{-0.2}^{+0.1}$~kpc from the joint sample. While all distance estimates are mutually consistent, the $V$-band estimate has significantly larger uncertainties due to the broader dust distributions for these echoes and fewer available images. Fig.~\ref{fig:joint} shows the 2D and explosion time-integrated posterior distributions for the joint sample. 


Fig.~\ref{fig:tyc_dist} also shows other estimates of the distance to Tycho. The two 21~cm H~\textsc{i} absorption
velocity distances of $1.7$-$3.7$~kpc \citep{albinson86} and $4.5\pm0.5$~kpc \citep{schwarz95}, illustrate the
problems with this method, driven in part by the velocity perturbations of a spiral arm. \cite{chevalier80} 
using a small number of proper motions and radial velocities found a distance of $2.3\pm0.5$~kpc. Using similar
methods, \citet{Smith1991} found a distance of $1.5$-$3.1$~kpc. \cite{hayato10}
combined X-ray proper motions from \citet{Katsuda2010} with their measurements of X-ray line widths to find a distance
of $4\pm1$~kpc. Based on the estimated peak brightness and light curve structure of the SN, \citet{krause08b}
estimated a distance of $3.5_{-0.9}^{+1.5}$~kpc and \citet{ruiz-lapuente04a} estimated a distance of $2.8\pm0.4$~kpc.
Finally, we note that \citet{volk08} argue that the lack of a $\gamma$-ray detection of Tycho requires a minimum
distance of $3.3$~kpc. Our results are generally consistent with these estimates if only because of their large uncertainties.

\begin{figure}

\begin{center}
    \includegraphics[width=.90\linewidth]{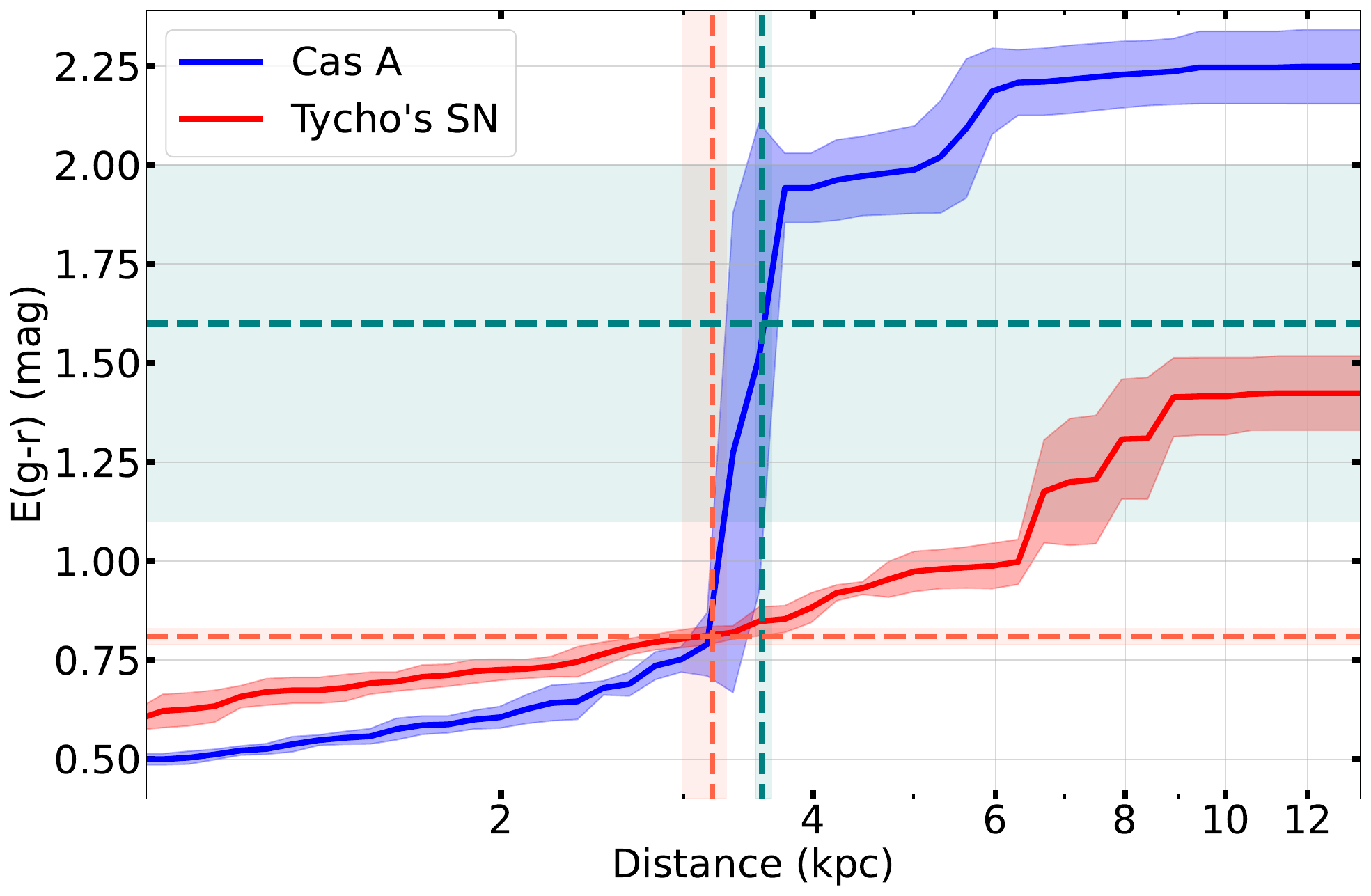}
    \caption{Average extinction as a function of distance at the explosion locations for each SN from the {\tt Bayestar19} reddening maps \citep{green19}. 
    Solid lines represent the extinction functions, and dashed lines represent the median distances found from the joint sample for each SN. Colored regions are the $1\sigma$ uncertainty calculated from the extinction map.}
    \label{fig:SNR_dust}
\end{center}
\end{figure}

Fig.~\ref{fig:SNR_dust} also shows the dust distribution along the line of sight to the center of Cas~A along
with our distance estimate. Unlike Cas~A, the dust is almost uniformly distributed in distance, and we find
a posterior estimate of the extinction towards the center of Tycho of $E(g-r)=0.82_{-0.02}^{+0.03}$. For
comparison, \citet{cassam07} find $N(H)\approx7\times10^{21} \mathrm{~cm~}^{-2}$ or $E(g-r)\simeq 0.8 \pm 0.1$ from {\it Chandra} X-ray observations
making the same assumptions about the uncertainties as for Cas~A. This agrees extremely well with our estimate.
At least for this sight line, the dust towards the SN would provide a much weaker distance estimate because the
gradient of extinction with distance is so weak. The estimate of the extinction from the column density would
need the precision of the uncertainties on the extinction posterior to produce a similar distance uncertainty.

\section{Discussion And Conclusions}
\label{sec:conclusion}

We present new distance measurements to the historical Cas~A and Tycho SNe by combining light echoes and 3D dust maps, finding fiducial distances of $3.6\pm0.1$ and $3.2_{-0.2}^{+0.1}$~kpc, respectively, that are most precise to date. The largest systematic
uncertainties arise from any miscalibrations in the dust distance scale, but as these are anchored on 
{\it Gaia} parallaxes, they are unlikely to be large. The uncertainties are driven by the low distance resolution of
the 3D dust maps and can be improved either by higher resolution maps or the addition of more echoes sampling different
line of sight dust distributions. The extinctions predicted by our distances agree with those expected from X-ray
measurements of the H~\textsc{i} column densities towards these two SNRs. Our method can be applied to any eruptive or explosive Galactic event luminous enough to produce resolved light echoes.

As noted in the introduction,
Tycho has been the focus of many searches for a surviving stellar binary
companion to the white dwarf which exploded (e.g., \citealt{RuizLapuente2004}, 
\citealt{Hernandez2009},
\citealt{Kerzendorf2013}, \citealt{Kerzendorf2018}). Finding
such a companion would be the first clear evidence of a single
degenerate channel for Type~Ia SNe, particularly given the failure to find evidence for the hydrogen expected to be stripped from the donor by the explosion in nebular phase spectra of large samples
of normal Type~Ia SNe (e.g., \citealt{tucker20}). 
The distance uncertainties for  both the candidate stars and Tycho itself
have made the search for an unbound companion challenging.
{\it Gaia} parallaxes have largely addressed the problem of 
the distances to the stars (see \citealt{RuizLapuente2019}),
and our echo distance largely solves the problem of the distance
to Tycho. \cite{RuizLapuente2019} listed 15 stars with \textit{Gaia}
distances consistent
with a distance of $2.7\pm 1.0$~kpc that roughly spans the earlier measurements in Fig.~\ref{fig:tyc_dist}. Only 7 of these stars have distances consistent
with our SNR distance, largely due to having larger parallax 
uncertainties. The occasionally favored candidate donor stars B and G are both clearly foreground stars.


Our ability to find light echoes with ASAS-SN slowly improves due to both the slowly increasing integration time and
the longer time baseline. With a sufficiently long baseline (nearly 20 years), we can search for echoes in the 
Magellanic Clouds where the long time scale is set by the expected time for an echo to traverse ASAS-SN's 16\arcsec\xspace
point spread function (PSF). The main problem is that older echoes have a lower surface brightness, and the high
stellar densities near the plane of the Galaxy limit ASAS-SN's surface brightness sensitivity
(at higher latitudes, its surface brightness sensitivity is quite good, reaching 27~mag/arcsec$^2$ in the g-band,
Jennerjahn et al. \textit{in prep}). Other existing surveys like ATLAS \citep{tonry18}, ZTF \citep{bellm19}, and the
Dragonfly array \citep{abraham14} should perform better at lower latitudes. The Vera C. Rubin Observatory \citep{ivezic19} can also contribute
to such searches at lower latitudes in the near future. Higher resolution also makes it easier to measure proper motions.
Effectively searching for echoes by eye continues to be a problem for all-sky rather than localized searches. Improved machine-learning techniques will likely be needed for such surveys.

\section*{Acknowledgements}

CSK and KZS are supported by NSF grants AST-1907570, 2307385, and 2407206. 
The Shappee group at the University of Hawai'i is supported with funds from NSF (grants AST-1908952, AST-1911074, \& AST-1920392) and NASA (grants HST-GO-17087, 80NSSC24K0521, 80NSSC24K0490, 80NSSC24K0508, 80NSSC23K0058, \& 80NSSC23K1431).
We thank Las Cumbres Observatory and its staff for their continued support of ASAS-SN. ASAS-SN is funded  by Gordon and Betty Moore Foundation grants GBMF5490 and GBMF10501 and  the Alfred P. Sloan Foundation grant G-2021-14192. 

Software used: \texttt{Astropy} (\citealt{astropy22}), {\sc Iraf} \citep[][]{tody86}, \texttt{Matplotlib} (\citealt{hunter07}), \texttt{NumPy} (\citealt{harris20}), \texttt{SciPy} \citep{virtanen20}.

\section*{Data Availability}
All data used in this paper are either public or in the tables associated with the paper.


\bibliographystyle{aasjournal}
\bibliography{echo}

\renewcommand{\arraystretch}{1.3}
\input{tables/Echoes}


\end{document}

%% file: defs.tex
\usepackage{xspace}

\newcommand{\asassn}{ASAS-SN\xspace}

\newcommand{\sbu}{mag~arcsec$^{-2}$\xspace}

\newcommand{\NH}{\hbox{\ensuremath{N(\rm H)}}\xspace}

\defcitealias{rest08b}{R08}

%% file: tables/Echoes.tex

\begin{table*}
\caption{Light Echoes}
\label{tab:LE_arc}
\centering
\begin{tabular}{cccccccccc}
\hline
SN$^a$ & ID$^b$ & Filter$^c$ & Group$^d$ & RA$^e$ & Dec$^e$ & MJD$^f$ & Min MJD$^g$ & Max MJD$^g$ & $\theta$$^h$ \\ 
\hline
Cas A & 1 & V & 2 & 00:20:04 & +61:58:61 & 57022 & 57008 & 57045 & 7.6508 \\ 
Cas A & 2 & V & 3 & 00:19:24 & +61:52:61 & 57022 & 57008 & 57045 & 7.5450 \\ 
Cas A & 3 & V & 4 & 00:19:44 & +61:44:61 & 57022 & 57008 & 57045 & 7.5445 \\ 
$\ldots$ & $\ldots$ & $\ldots$ & $\ldots$ & $\ldots$ & $\ldots$ & $\ldots$ & $\ldots$ & $\ldots$ & $\ldots$ \\ 
Tycho & 1 & V & 1 & 00:49:19 & +58:43:04 & 57026 & 57008 & 57045 & 6.126 \\ 
Tycho & 2 & V & 3 & 01:06:38 & +59:21:11 & 57026 & 57008 & 57045 & 6.825 \\ 
Tycho & 3 & V & 1 & 00:49:23 & +58:42:49 & 57326 & 57194 & 57413 & 6.134 \\ 
$\ldots$ & $\ldots$ & $\ldots$ & $\ldots$ & $\ldots$ & $\ldots$ & $\ldots$ & $\ldots$ & $\ldots$ & $\ldots$ \\ 
 
\hline
\end{tabular}
\smallskip
\\
\raggedright
\noindent This table is available in its entirety in a machine-readable form in the online journal. A portion is shown here for guidance regarding its form and content.\\
$^a$ SN or SN remnant associated with the echo.\\
$^b$ Identification number associated with the echo per SN.\\
$^c$ Filter associated with each echo.\\
$^d$ Group number associated with each echo.\\
$^e$ Right ascension and declination for the J2000 epoch. \\
$^f$ Averaged Modified Julian Date of the observation.\\
$^g$ Minimum and maximum Modified Julian Date of observations \\
$^h$ Separation of the echo from the center of the SN in degrees.\\
\end{table*}